\documentclass[showpacs,preprintnumbers,amsmath,
amssymb,overcite]{revtex4}
\usepackage{graphicx}
\usepackage{dcolumn}
\usepackage{bm}

\begin{document}

\title{Analysis of the consistency of kaon photoproduction data 
with $\Lambda$ in the final state}

\author{P. Byd\v{z}ovsk\'y\footnote{e-mail: bydz@ujf.cas.cz}}
\affiliation{Nuclear Physics Institute, \v{R}e\v{z} near Prague, 
Czech Republic}
\author{T. Mart}
\affiliation{Departemen Fisika, FMIPA, Universitas Indonesia, 
Depok 16424, Indonesia}

\date{\today}

\begin{abstract}
The recent CLAS 2005, SAPHIR 2003, LEPS, and the old, pre-1972, 
data on $K^+\Lambda$ photoproduction are compared with theoretical 
calculations in the energy region of $E_{\gamma}^{\rm lab}<$ 2.6 GeV 
in order to learn about their mutual consistency. The isobaric models 
Kaon-Maid and Saclay-Lyon, along with new fits to the CLAS data are 
utilized in this analysis. The SAPHIR 2003 data are shown to be 
coherently shifted down with respect to the CLAS, LEPS, and pre-1972 
data, especially at forward kaon angles. The CLAS, LEPS, and pre-1972 
data in the forward hemisphere can be described satisfactorily by using 
the isobaric model without hadronic form factors. The inclusion of 
the hadronic form factors yields a strong suppression of the cross 
sections at small kaon angles and c.m. energies larger than 1.9 GeV, 
which is not observed in the existing experimental data. We demonstrate 
that the discrepancy between the CLAS and SAPHIR data has a significant 
impact on the predicted values of the mass and width of 
the ``missing-resonance'' $D_{13}(1895)$ in the Kaon-Maid model.

\end{abstract}

\maketitle


\section{Introduction}
Kaon photoproduction on the nucleon provides an important tool 
for understanding the dynamics of hyperon-nucleon systems.
Accurate information on the elementary amplitude is vital for calculating  
the cross sections of the hypernuclear photoproduction, since 
the amplitude serves as the basic input, which determines the 
accuracy of predictions \cite{Motoba,ProdH}. At present, these 
calculations can be compared  with high resolution spectroscopy 
data of the hypernuclei, which are available from the experiments 
performed at the Jefferson Laboratory \cite{Hashimoto}. 
Since the hypernucleus production cross section is sensitive 
to the elementary amplitude, especially at forward kaon angles, 
a precise description of the elementary process at this
kinematics is obviously desired.

\begin{figure}[htb]
\includegraphics[width=.55\textwidth,angle=-90]{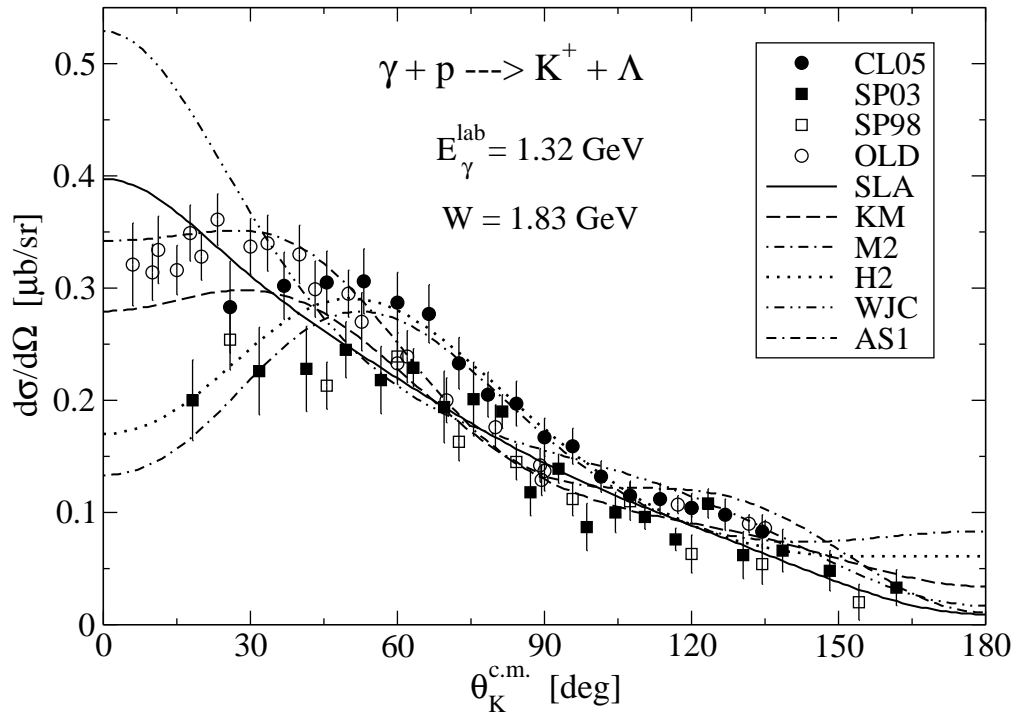}
\caption{Comparison of various data sets with predictions of 
different phenomenological models,
Saclay-Lyon A (SLA) \cite{SLA98}, Kaon-Maid (KM) \cite{Ben99},
M2, H2 \cite{HYP03}, Williams-Ji-Cotanch (WJC) \cite{WJC92}, and
Adelseck-Saghai (AS1) \cite{AS90}. Data are adopted from 
Refs.~\cite{CL05}(CL05), \cite{SP03}(SP03),
\cite{SP98}(SP98), and \cite{AS90}(OLD).
Total error bars are indicated in the plot.}
\label{crs_comp}
\end{figure}

The two sets of ample, good quality, experimental data provided
recently by the CLAS (CL05) \cite{CL05} and SAPHIR (SP03) \cite{SP03} 
collaborations were expected 
to help us learn more about the process; however, they reveal a lack 
of consistency at forward and backward kaon angles \cite{CL05} 
(see also Ref.~\cite{HYP03} in which results of the first analysis 
of the CLAS data \cite{oldCLAS} were used). 
The previous SAPHIR data by Tran {\it et al.} (SP98) \cite{SP98} 
also display different behavior at small kaon angles compared to 
that observed in the old pre-1972 data, e.g., 
from Bleckmann {\it et al.} \cite{Bleckmann}
(hereafter referred to as OLD). 
The uncertainty in the experimental information causes a wide
range of model predictions at forward kaon angles. The situation is
illustrated in Fig.~\ref{crs_comp}, where the CL05, SP03, SP98 and
OLD data (as listed in Ref.~\cite{AS90}) are compared 
with predictions of different phenomenological models. Obviously, 
the data and the models, which were fitted to various data sets, 
differ significantly for $\theta_K < 45^{\circ}$, which leads to a 
large input uncertainty in the hypernuclear calculations \cite{ProdH}.
%
%

At present, there are two large data sets, the latest CLAS and
SAPHIR ones, with comparable statistical significance, but
they diverge in some kinematic regions. Measurements of the 
differential cross sections at small kaon angles from 
LEPS \cite{LEPS} provide another good quality data set for 
energies from 1.5 to 2.4 GeV. These data are consistent with 
the CLAS but not the SAPHIR data. The older data, SP98 and OLD, 
are scarce; and  for $\theta_K < 45^\circ$, they also reveal some 
discrepancies, as shown by open squares \cite{SP98} and open 
circles \cite{Bleckmann} in Fig.\,\ref{crs_comp}. This situation 
clearly indicates that before a reliable determination of 
the parameters of a model for the elementary process can be 
performed, we have to decide which data sets 
are consistent with each other and which can thus 
be used in fitting the models. 
The purpose of this work is to analyze 
the mutual consistency and similarities of the data sets by using 
selected isobaric models. The analysis will enable a better 
determination of the elementary amplitude, especially at forward 
angles. We also discuss certain problems of the isobaric models 
with the description of the data at forward directions.

This paper is organized as follows:
In Sec.~\ref{analysis}, the basic formalism and definitions of the 
kinematic regions used in this analysis are given. 
The experimental data and the utilized models are briefly discussed 
in Secs. II A and II B, respectively. In Sec.~\ref{Results_and_discussion}, 
results are presented and discussed. Conclusions are given in 
Sec.~\ref{Conclusions}.

\section{Analysis}
\label{analysis}
Although there are some kinematics overlaps of the considered data
sets, an interpolation by using an analytical formula is still necessary 
to perform a direct comparison. To avoid this we compare
the observed cross sections with predictions of theoretical models. 
For this purpose, we calculate the relative deviation for each data 
point as done in the analysis of OLD data~\cite{AS90}, 
\begin{equation}
R_i=\frac{\sigma_i^{\rm exp}-\sigma^{\rm th}(E_i,\theta_i)}
{\Delta\sigma_i^{\rm stat}},
\label{deviation}
\end{equation}
where $\sigma_i^{\rm exp}$ and $\Delta\sigma_i^{\rm stat}$ are  
the measured value and its statistical uncertainty, respectively, 
at the kinematics given by the photon laboratory energy $E_i$
and the kaon center of mass angle $\theta_i$. The theoretical 
value $\sigma^{\rm th}(E_i,\theta_i)$ is calculated within  
a particular isobaric model at the appropriate kinematic point. 
If the theoretical values correctly describe the reality and 
the experimental values are randomly scattered around them 
with the variance given by $\Delta\sigma_i^{\rm stat}$, then 
the variable $R_i$ possesses a normal distribution with the mean 
$\mu$=0 and the variance $\sigma^2$=1. We are, however, far from 
this ideal case. The distribution of $R_i$, calculated 
for a particular model and experimental data set, which clearly 
depends on the chosen model, thus, characterizes a consistency of 
the model with the data set. To this end, we also calculate 
the required parameters of the distribution, i.e., the mean value
\begin{equation}
\langle R\rangle = \frac{1}{N}\sum_{i=1}^{N}R_i,
\end{equation}
the second algebraic moment
\begin{equation}
\langle R^2\rangle = \frac{1}{N}\sum_{i=1}^{N}R_i^2 = \chi^2 / N,
\label{chi2}
\end{equation}
the standard deviation
\begin{equation}
s^2 = \frac{N}{N-1}\langle(\Delta R)^2\rangle  = 
\frac{N}{N-1}(\langle R^2\rangle -\langle R\rangle ^2),
\label{stdev}
\end{equation}
and the number of data points with $R_i$ in the interval of
($\langle R\rangle-2$, $\langle R\rangle+2$) relative to the number
of data $N$, which is denoted by $N_2$ (in \%). The summations run 
over the data points included in the sample. The agreement between 
model predictions and experimental data is expressed by $\chi^2/N$ 
which includes also information on the data dispersion. The mean 
value $\langle R\rangle$ shows a coherent shift of the data with 
respect to the model predictions. The condition $\langle R\rangle=0$ 
is necessary for the model and data to describe simultaneously the 
reality (a population). 

Provided that the data are randomly scattered around the theoretical
values $\sigma^{\rm th}(E_i,\theta_i)$ with the variance
$\Delta\sigma_i^{\rm stat}$, i.e. \{$R_i$, i=1, $N$\} is a random sample 
with a normal distribution, the hypothesis that the true value of 
the mean $\langle R\rangle$ equals zero (the null hypothesis) can 
be tested by calculating the statistical parameter (Student's 
t-variable) \cite{StatMan}:
\begin{equation}
z_1 = \sqrt{N-1}\frac{\langle R\rangle}{\sqrt{\langle(\Delta R)^2\rangle}} .
\label{z_1stat}
\end{equation}
Here, the variance of the normal distribution of $R_i$ is supposed
to be known and can be approximated by the standard deviation (\ref{stdev}),
since $N$ is sufficiently large ($>30$) for the assumed data sets.
The hypothesis will be rejected with a confidence level of $\alpha$
if $|z_1|>z_{\alpha/2}$, where the critical value $z_{\alpha/2}$ = 1.96
and 2.58 for the confidence level of 5\% and 1\%,
respectively \cite{StatMan}.

In this analysis, we define two types of data samples taken from 
each of the experimental data sets with different kinematics, i.e.:
\begin{itemize}
\item sample A: $0.91$ GeV$ < E_i < 2.6$ GeV 
and $0^\circ < \theta_i < 180^\circ$,
\item sample B: $0.91$ GeV$ < E_i < 2.6$ GeV 
and $0^\circ < \theta_i < 60^\circ$.
\end{itemize}
The statistics of sample B are more sensitive to the differences 
between the data and model predictions at forward angles, where 
the largest discrepancies among the data sets and models exist 
(see Fig.~\ref{crs_comp}). Polarization and total cross section data 
are not considered in our analysis.

\subsection {Experimental data}
\label{expdata}
The following experimental data sets consisting of differential cross 
sections have been used in calculating $R_i$:
\begin{itemize}
\item the CLAS data \cite{CL05}, labeled by CL05 in the figures and tables,
\item the latest SAPHIR data \cite{SP03} (SP03),
\item the LEPS data \cite{LEPS} (LEPS), and
\item the set of pre-1972 data (OLD), used in the analysis of Adelseck 
and Saghai~\cite{AS90}.
\end{itemize}
Note that the last set is listed in Table IX of Ref.~\cite{AS90}, 
except for the data by Decamp {\it et al.} (Orsay data). In the CL05 
data set, we only consider the data points from threshold up to 
$E_{\gamma}^{\rm lab}$ = 2.6 GeV ($W=2.4$ GeV, see samples A 
and B), in order to make an overlap with the SP03 data set and to 
maintain a reasonable description of the cross sections provided 
by isobaric models. 

The statistical uncertainties of the cross sections were used in the
analysis and in the fits of the new models (see the next subsection).
The systematic uncertainty of CL05 was estimated to be 8\%
except for the forward-most angle bins, where the uncertainty amounts
to 11\% \cite{CL05}. For the SP03 \cite{SP03} and OLD \cite{AS90} 
data the systematic error bars were reported for each data point. 
The overall systematic uncertainty of the LEPS data was estimated 
to be 7\% \cite{LEPS}.

It was shown that the LEPS data are in good agreement with the CLAS data 
within the total uncertainty and are systematically higher than the SP03 
data at all angles ($\theta_{\rm K}^{\rm c.m.} < 41^\circ$) \cite{LEPS}. 
The SP03 data are systematically smaller than the CL05 ones for $W>1.75$~GeV. 
We note that an energy-independent scale factor of about 3/4 between the 
CL05 and SP03 results was suggested in Ref.~\cite{CL05}.  

\subsection{Models used in the analysis}
\label{models}

Theoretical values of the cross sections in Eq.~(\ref{deviation})
were calculated within the isobaric models for the photoproduction
of $K^+$ on the proton. In these models the amplitude is
constructed by using the Feynman diagrammatic technique, assuming
only contributions of the tree-level diagrams. The effective
Lagrangian is written in terms of resonant states and asymptotic 
particles. Because of the absence of a dominant resonance, as in the 
case of pion and $\eta$ photoproductions, various nucleon and hyperon 
resonances are considered, which results in a copious number of 
models \cite{Byd03}. Hadrons were 
supposed to be pointlike particles in the strong vertices in some 
models \cite{SL96,WJC92,AS90,SLA98} but, in the newest 
ones \cite{Ben99,HYP03,Jan01}, the hadron structure is considered  
by means of hadronic form factors. The effective coupling constants 
in the models were determined by fitting the appropriate observables 
to experimental data.

In our analysis, the Saclay-Lyon (SL) \cite{SL96} and Kaon-Maid
(KM) \cite{Ben99} models were adopted. Common to these models is that, 
besides the extended Born diagrams, they also include kaon resonances
$K^*(890)$ and $K_1(1270)$. In Ref. \cite{WJC92}, it was shown that 
these $t$-channel resonant terms together with the nucleon ($s$-channel) 
and hyperon ($u$-channel) resonances can improve the agreement with 
the experimental data in the intermediate energy region. The models 
differ in the choice of the particular $s$- and $u$-channel 
resonances in the intermediate state, in the treatment of the hadron 
structure, and in the set of experimental data to which the free 
parameters were adjusted. However, the two main coupling constants,
$g_{KN\Lambda}$ and $g_{KN\Sigma}$, fulfill the limits of 20\% broken
SU(3) symmetry \cite{SL96} in both models.

In the SL model, four hyperon and three nucleon resonances with 
the spin up to 5/2 are included and their coupling constants were 
fitted to the OLD data set \cite{AS90} and the first results of SAPHIR 
by Bockhorst {\it et al.} \cite{SAPHIR94}. In the KM model, four 
nucleon but no hyperon resonances were assumed and the parameters 
of the model were fitted to the OLD and SP98 \cite{SP98} data sets.
The SL and KM models were expected to provide reasonable results for 
photon energies below 2.2 GeV. In our analysis, however, we consider 
the results of these models for energies up to 2.6 GeV.

In the SL model, hadrons are treated as pointlike objects, in contrast 
to the KM model in which hadronic form factors (h.f.f.) are inserted 
in the hadronic vertices \cite{Ben99}. The inclusion of h.f.f. in the 
isobaric model substantially improves the agreement with the higher 
energy data. However, it appears to be the source of the significant 
suppression of the cross sections at very small kaon angles and higher 
energies ($E_{\gamma} > 1.7$ GeV, see Fig.~4a in Ref.~\cite{ProdH} and 
Fig.~\ref{crs_comp} for M2 and H2 models, which include h.f.f. and 
were fitted to the results of the first analysis of the CLAS 
data \cite{HYP03}). 
 
In addition to the KM and SL models, we have also included two new 
models, which are referred to as fit 1 and fit 2. Fit 1 includes, 
besides the Born terms and kaon resonances $K^*$ and $K_1$, the same 
$s$-channel resonances as in the KM model: $S_{11}(1650), P_{11}(1710), 
P_{13}(1720)$ and $D_{13}(1895)$. The latter is known as 
the ``missing'' resonance, a resonance predicted by the quark model but 
not yet listed in the Particle Data Book \cite{Ben99}. 
Its presence in the model of this type is, however, important for 
the description of the resonant structure seen in the SAPHIR and 
CLAS data \cite{Ben99,HYP03}. The background part of the amplitude 
is improved by assuming the $u$-channel resonances as suggested 
by Janssen {\it et al.} \cite{Jan01}. Particularly, 
$S_{01}(1670)$ and $P_{11}(1660)$ hyperon states were chosen in 
fit 1, as they give the best agreement with the data. The hadron 
structure in the strong vertices is modeled by the dipole-type 
form factors introduced by a certain gauge-invariant 
technique \cite{DW01}. The cutoff parameters in the form factors 
of the Born and resonant contributions are independent. The free 
parameters of fit 1, i.e. the coupling constants and cutoffs, were 
determined by fitting the differential cross sections to all 
CLAS data in the energy region of $E_{\gamma}^{\rm lab} < 2.6$ GeV 
(see the definition of sample A in Section~\ref{analysis}).

The model fit 1 exhibits a strong suppression of the cross sections at 
small kaon angles for $E_{\gamma} >1.5$~GeV as discussed above in 
connection with h.f.f. This pattern, being connected with a strong 
suppression of the Born terms, particularly the electric part of the 
proton exchange, causes large deviations of the model predictions 
from the data at small angles, which precludes analysis of the data 
at forward angles. 
To have a more realistic description of the forward-angle data we 
assume also a model \underline{without h.f.f.}, fit 2. 
The resonance content of fit 2 was motivated by the SL model, which  
shows a better agreement with the data in the forward hemisphere 
than the KM model, especially for energies $E_{\gamma} > 1.7$~GeV 
($W > 2$~GeV, see the next section). Therefore, the following 
resonances were included in fit 2: 
the $t$ channel, $K^*$ and $K_1$; $s$ channel, $P_{13}(1720)$, 
$D_{15}(1675)$, and $D_{13}(1895)$; and  $u$ channel, $S_{01}(1405)$,  
$S_{01}(1670)$, and $P_{01}(1810)$. 
The nucleon $P_{11}(1440)$ resonance, which was included in SL but 
whose coupling constant is very small \cite{SL96}, was repalced by 
$D_{13}(1895)$ to better describe the resonance behavior of the data.
The presence of the hyperon $P_{11}(1660)$ resonance, which was also included 
in the SL model, appears to be irrelevant in the forward-angle region. 
On the other hand, the higher spin (5/2) s-channel resonance  
$D_{15}(1675)$ appears to be very important for reduction of the cross 
section at energy $W>1.8$~GeV and forward angles. Its coupling constants 
appear to be much larger than those of the other $s$-channel 
resonances in fit 2. Parameters of fit 2 were fitted to CL05 for energy 
up to 2.6 GeV but for $\theta_{\rm K}^{\rm c.m.} < 90^\circ$. 
Note that the kaon angles were limited in order to avoid problems 
of these models at backward regions \cite{Byd03} (see also the next section) 
and to achieve a good agreement with the data at forward angles. 

In both fits, statistical uncertainties of experimental data (see 
Sect.~\ref{expdata}) were taken into account and the two main 
coupling constants were forced to keep the limits of 20\% broken SU(3) 
symmetry:
$-4.4 \leq g_{KN\Lambda}/\sqrt{4\pi} \leq -3.0$ and
$0.8 \leq g_{KN\Sigma}/\sqrt{4\pi} \leq 1.3$. The values of the 
cutoff parameters were also confined in the range of 
$0.6$ GeV $\leq \Lambda \leq 2.0$ GeV. The best values of 
$\chi^2/{\rm n.d.f.}$ for fit 1 and fit 2 are 3.46 and 1.80, respectively.

\section{Results and discussion}
\label{Results_and_discussion}

\begin{figure}[!ht]
\includegraphics[width=.85\textwidth]{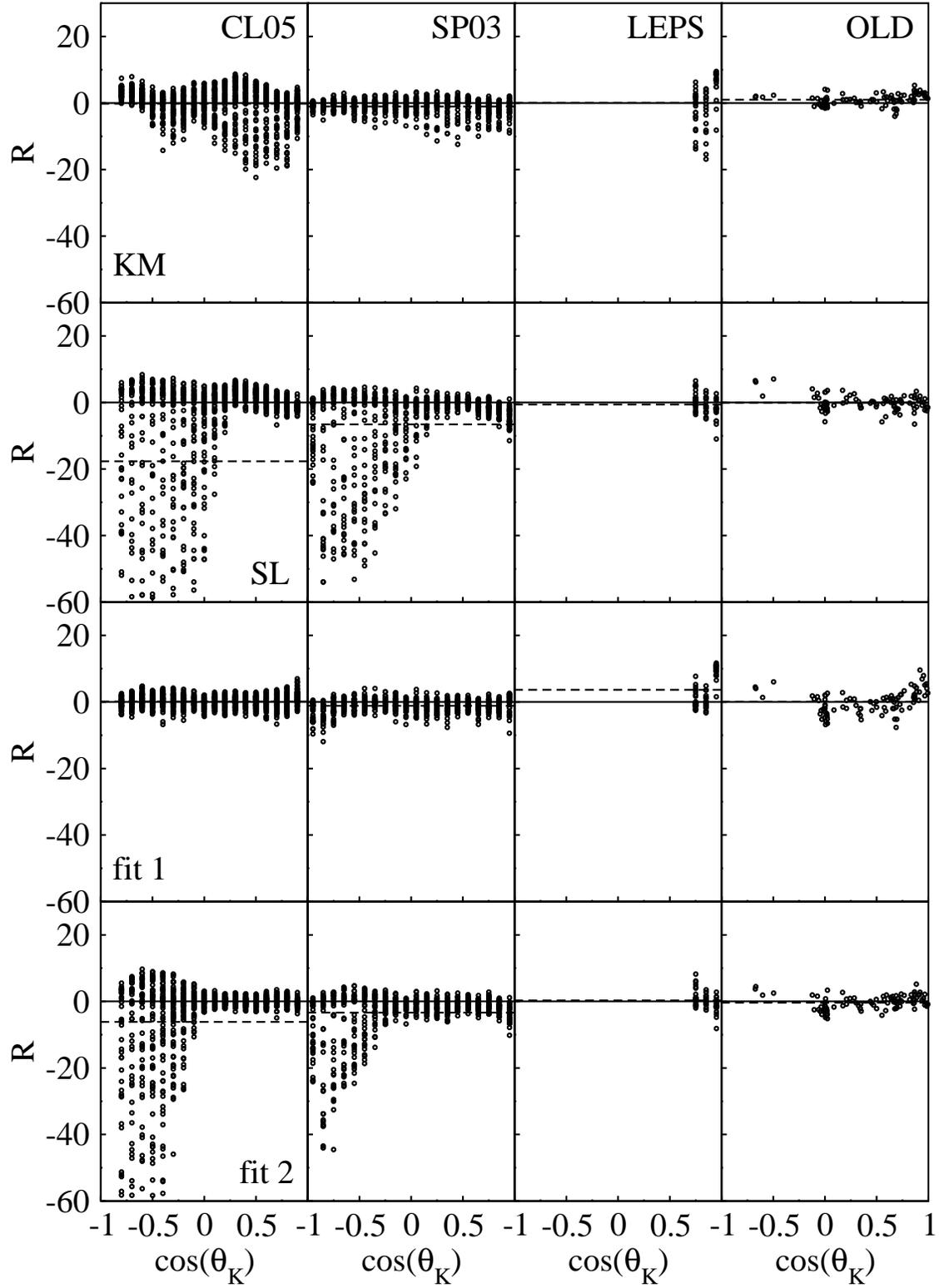}
\caption{Deviations of the experimental data points from the 
predictions of the models as a function of cosine of the kaon c.m. 
angle. The data for the photon laboratory energy below 2.6 GeV 
are assumed. The mean values of $R$ are represented by dashed 
lines. Panels in one row correspond to the same theoretical 
model, whereas panels in the same column use the same 
experimental data set.}
\label{anal_2_E}
\end{figure}

\begin{figure}[!ht]
\includegraphics[width=.85\textwidth]{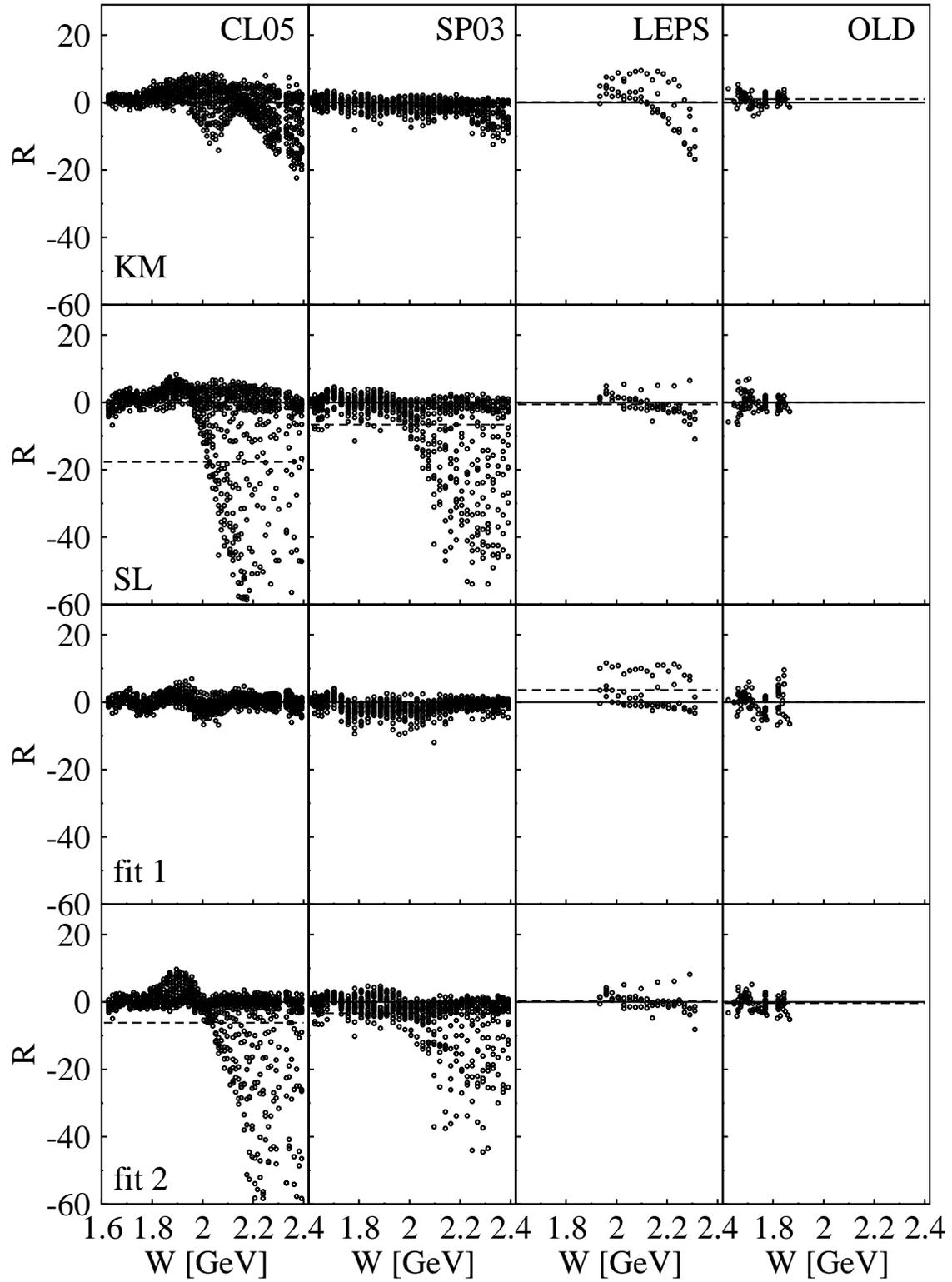}
\caption{As in Fig.~\ref{anal_2_E}, but the deviations are 
a function of the total c.m. energy. The data cover the full range 
of the kaon c.m. angle (sample A).}
\label{anal_1_E}
\end{figure}

\begin{figure}[!ht]
\includegraphics[width=.85\textwidth]{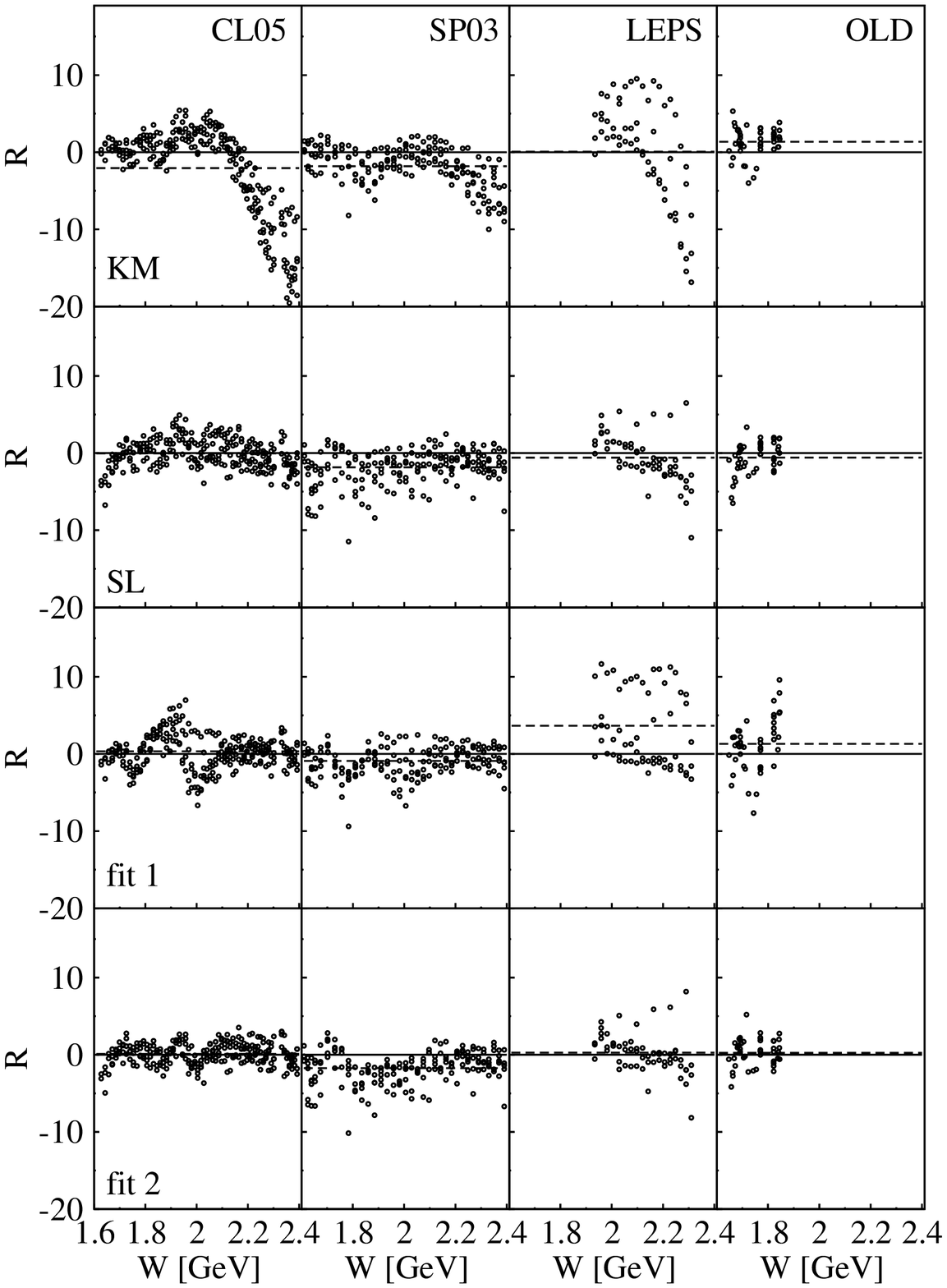}
\caption{As in Fig.~\ref{anal_1_E}, but for experimental data with
$0^\circ <\theta_{\rm K}^{\rm c.m.} < 60^\circ$ (sample B).}
\label{anal_1_F}
\end{figure}

The statistical parameters of the distributions defined in
Section~\ref{analysis} for samples A and B are listed in 
Tables~\ref{sampleA} and \ref{sampleB}, respectively, while 
the relative deviations of the experimental values from 
theoretical predictions, $R_i$, are displayed in 
Figs.~\ref{anal_2_E}-\ref{anal_1_F}. The corresponding mean 
values, $\langle R\rangle$, are indicated by the dashed lines 
in each panel of the figures. Panels in a row correspond to 
the particular model, whereas panels in a column use
the same experimental data set (see Section~\ref{expdata} for the
definitions of the data labels and Section~\ref{models} for the
definitions of the model labels). In Figs.~\ref{anal_2_E} and
\ref{anal_1_E}, the deviations of each data point for all kaon
angles (sample A) are plotted as functions of the kaon 
c.m. angle and the total c.m. energy, respectively. 
Figure~\ref{anal_1_F} shows results for the forward angles, 
from $0^\circ$ up to $60^\circ$ (sample B).

\renewcommand{\baselinestretch}{1}
%
%
\begin{table}[t]
\caption{Statistical parameters of sample A. \label{sampleA}}
\begin{center}
\begin{ruledtabular}
\begin{tabular}{lrrrrrr}
\multicolumn{1}{c}{data set} & \multicolumn{1}{c}{$N$} &
\multicolumn{1}{c}{$\langle R\rangle$} &
\multicolumn{1}{c}{$\chi^2/N$} &
\multicolumn{1}{c}{$\langle (\Delta R)^2\rangle$} &
\multicolumn{1}{c}{$z_1$} & \multicolumn{1}{c}{$N_2$} (\%)\\
\hline
\multicolumn{7}{c}{model KM}\\
CL05  & 1109 & -0.22 &  25.7 &  25.7\ \ \ \  &  -1.41 &  37.1\\
SP03  &  701 & -1.04 &  6.69 &  5.60\ \ \ \  &  -11.7 &  68.9\\
LEPS  &   60 &  0.08 &  45.4 &  45.4\ \ \ \  &   0.09 &  26.7\\
OLD   &   91 &  1.00 &  3.82 &  2.82\ \ \ \  &   5.66 &  74.7\\
\hline
\multicolumn{7}{c}{model SL}\\
CL05  & 1109 &  -17.7 &  2145 &  1832\ \ \ \  &  -13.7 &   1.5\\
SP03  &  701 &  -6.59 &   198 &   155\ \ \ \  &  -14.0 &   7.0\\
LEPS  &   60 &  -0.60 &  10.1 &  9.70\ \ \ \  &  -1.47 &  51.7\\
OLD   &   91 &  -0.09 &  5.72 &  5.71\ \ \ \  &  -0.35 &  68.1\\
\hline
\multicolumn{7}{c}{fit 1 }\\
CL05  & 1109 &  0.15 &  3.42 &  3.39\ \ \ \  &   2.66 &  72.8\\
SP03  &  701 & -1.24 &  5.89 &  4.36\ \ \ \  &  -15.7 &  70.6\\
LEPS  &   60 &  2.96 &  31.5 &  22.7\ \ \ \  &   4.76 &  20.0\\
OLD   &   91 & -0.04 &  11.6 &  11.6\ \ \ \  &  -0.10 &  47.3\\
\hline
\multicolumn{7}{c}{fit 2}\\
CL05  & 1109 & -6.81 &   544 &   498\ \ \ \  &  -10.2 &   2.5\\
SP03  &  701 & -3.61 &  66.8 &  53.8\ \ \ \  &  -13.0 &  30.1\\
LEPS  &   60 &  0.26 &  6.76 &  6.69\ \ \ \  &   0.77 &  70.0\\
OLD   &   91 & -0.32 &  4.83 &  4.72\ \ \ \  &  -1.41 &  59.3\\
\end{tabular}
\end{ruledtabular}
\end{center}
\end{table}

%
%
\begin{table}[t]
\caption{Statistical parameters of sample B.\label{sampleB}}
\begin{center}
\begin{ruledtabular}
\begin{tabular}{lrrrrrr}
\multicolumn{1}{c}{data set} & \multicolumn{1}{c}{$N$} &
\multicolumn{1}{c}{$\langle R\rangle$} &
\multicolumn{1}{c}{$\chi^2/N$} &
\multicolumn{1}{c}{$\langle (\Delta R)^2\rangle$} &
\multicolumn{1}{c}{$z_1$} & \multicolumn{1}{c}{$N_2$} (\%)\\
\hline
\multicolumn{7}{c}{model KM}\\
CL05  & 252 & -2.06 &  37.2 &  33.0\ \ \ \  &  -5.67 &  17.1\\
SP03  & 178 & -1.82 &  10.0 &  6.75\ \ \ \  &  -9.30 &  53.4\\
LEPS  &  60 &  0.08 &  45.4 &  45.4\ \ \ \  &   0.09 &  26.7\\
OLD   &  46 &  1.35 &  5.43 &  3.60\ \ \ \  &   4.78 &  73.9\\
\hline
\multicolumn{7}{c}{model SL}\\
CL05  & 252 & -0.05 &  3.69 &  3.68\ \ \ \  &  -0.37 &  70.2\\
SP03  & 178 & -1.84 &  8.87 &  5.48\ \ \ \  &  -10.5 &  65.2\\
LEPS  &  60 & -0.60 &  10.1 &  9.70\ \ \ \  &  -1.47 &  51.7\\
OLD   &  46 & -0.59 &  4.60 &  4.25\ \ \ \  &  -1.91 &  73.9\\
\hline
\multicolumn{7}{c}{fit 1}\\
CL05  & 252 &  0.22 &  4.91 &  4.86\ \ \ \  &   1.60 &  65.1\\
SP03  & 178 & -1.00 &  4.49 &  3.50\ \ \ \  &  -7.08 &  75.8\\
LEPS  &  60 &  2.96 &  31.5 &  22.7\ \ \ \  &   4.76 &  20.0\\
OLD   &  46 &  1.12 &  12.7 &  11.4\ \ \ \  &   2.23 &  52.2\\
\hline
\multicolumn{7}{c}{fit 2}\\
CL05  & 252 &  0.11 &  1.98 &  1.97\ \ \ \  &   1.25 &  84.5\\
SP03  & 178 & -1.70 &  7.37 &  4.47\ \ \ \  &  -10.7 &  67.4\\
LEPS  &  60 &  0.26 &  6.76 &  6.69\ \ \ \  &   0.77 &  70.0\\
OLD   &  46 &  0.37 &  3.23 &  3.09\ \ \ \  &   1.42 &  78.3\\
\end{tabular}
\end{ruledtabular}
\end{center}
\end{table}

%
%
%
%
%
%

Table~\ref{sampleA} reveals that the $\chi^2/N$ values of the model KM are 
much larger for the CL05 and LEPS data sets than for the SP03 and OLD ones. 
The SP03 data seem also to be scattered closer to the model predictions 
than the CL05 and LEPS as indicated by the values of $N_2$. However, for 
sample A the average relative statistical uncertainties, 
$\Delta \sigma^{\rm stat}/\sigma^{\rm exp}$, 
are smaller for the CL05 (10\%) and LEPS (6\%) than for the SP03 (38\%) data, 
which makes the values of $R_i$, and therefore $\chi^2/N$, much 
smaller for the SP03. The statistics $|z_1|$ in Table~\ref{sampleA}, 
which is not sensitive to this effect, shows that the KM model provides 
a good description of the CLAS ($|z_1|=1.41$) and LEPS ($|z_1|=0.09)$ 
data sets (in this case, if we reject the null hypothesis, there is 
a large probability that we are wrong). On the contrary, the KM model does 
not seem to be consistent with the SP03, i.e., $|z_1|=11.7 \gg 2.58$ for 
the confidence level of 1\% (the null hypothesis can be safely rejected).

The very large values of  $\chi^2/N$ and $|\langle R\rangle|$ for 
the model SL with CL05 and SP03 data in comparison with those for 
the KM model are mainly 
due to the deficiency of the SL model in describing the data at backward 
angles ($\theta_{\rm K}^{\rm c.m.} > 100^\circ$) for $W>2$ GeV, as can be 
clearly seen in Figs.~\ref{anal_2_E} and \ref{anal_1_E}. 
However, at forward angles, the SL model gives a better agreement with 
the CL05 and OLD data than the KM model [see the statistics 
$\langle R\rangle$, $\chi^2/N$, 
$z_1$, and $N_2$ in Table~\ref{sampleB} (sample B) and 
Fig.~\ref{anal_1_F}]. This indicates that the SL model (without 
h.f.f.) is more suited for the description of the forward-angle data 
than KM. The SL model also agrees well with the OLD data at backward 
angles (Table~\ref{sampleA}), since these data are limited to  
the photon energies up to 1.5 GeV, and, moreover, they were used 
to fit the parameters of the model.

The new model, fit 1, which was fitted to the CL05 data for all angle bins 
(sample A), gives small $\langle R\rangle$ (0.15) but quite large 
$z_1$ (2.66) for CL05 (Table~\ref{sampleA}), which suggests that 
the model describes the data with a confidence level smaller than 1\%. 
The largest deviations $R_i$ are found, however, for the data at small angles 
and in the energy range of 1.8 -- 2 GeV (see Fig.~\ref{anal_1_F}). 
Comparison of the $\langle R\rangle$ and $\chi^2/N$  for fit 1 with CL05 
in Tables~\ref{sampleA} and \ref{sampleB} also indicates that the 
model systematically underpredicts the data for small angles. 
The underprediction of the most-forward-angle cross sections by fit 1 
is also apparent for the LEPS data, as obviously shown by 
Figs.~\ref{anal_2_E} -- \ref{anal_1_F}. 
%
%
\begin{figure}[!ht]
\includegraphics[width=.5\textwidth,angle=270]{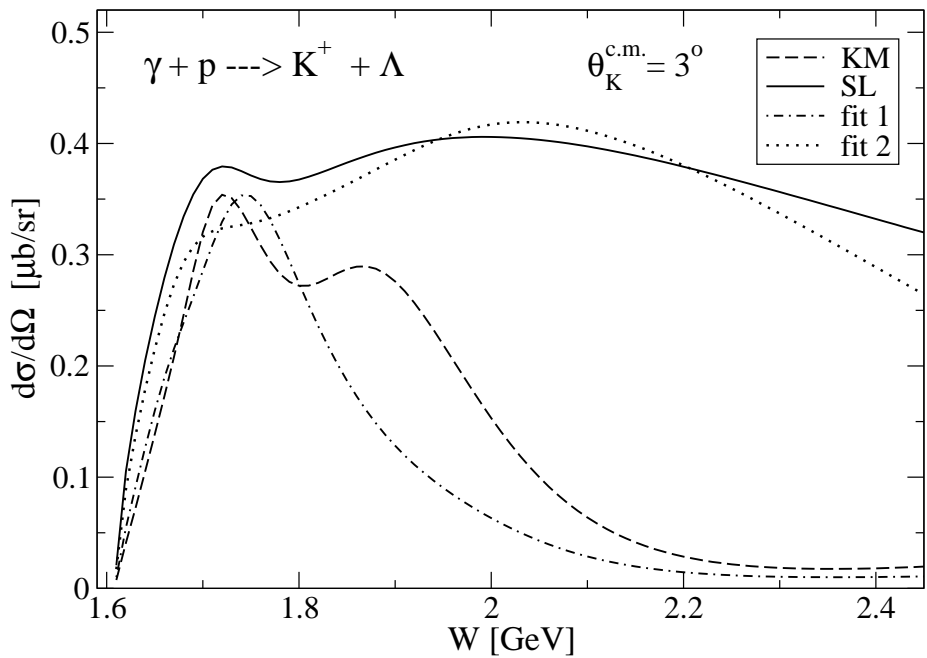}
\caption{Cross sections at $\theta_{\rm K}^{\rm c.m.}=3^\circ$ as 
predicted by several isobaric models (see text for details).}
\label{theta3}
\end{figure}
In Fig.~\ref{theta3} we demonstrate the behavior of the forward-angle 
($\theta_{\rm K}^{\rm c.m.}=3^\circ$) cross sections as a function of 
energy for the assumed models. The cross section suppression 
predicted by the models with h.f.f. (KM and fit 1) is clearly seen 
for $W>1.8$ GeV. These results suggest that models with h.f.f. 
introduced in a certain way \cite{DW01} cannot provide a realistic 
description of the forward-angle data. Therefore, the concept of 
h.f.f. \cite{Ben99,DW01} should be further investigated to correct 
the too strong damping of the cross sections at forward angles and 
larger energies, which is not observed in the existing data.

As expected, 
the results of fit 2 (fitted only to the forward-hemisphere data)  
with CL05 at forward angles (sample B) are very good (see 
Table~\ref{sampleB} for the statistics). However, at backward angles, 
fit 2 reveals the same deficiency as seen with the SL model (see 
Figs.~\ref{anal_2_E} and \ref{anal_1_E} and Table~\ref{sampleA}), 
although in general fit 2 is much better. 
The fit 2 model also provides better statistics at forward angles 
for the LEPS and OLD data than for SP03 (see Table~\ref{sampleB}), 
which quantitatively demonstrates that at forward angles the CL05, 
LEPS, and OLD data can be described simultaneously  by  
an isobaric model without h.f.f. Most of the data are scattered 
near the model predictions as shown by the large values of $N_2$ 
(defined with the statistical uncertainty) and 
$|\langle R\rangle|\approx 0$. 
The values of $|z_1|$ are small enough in comparison with the value 
for the 5\% confidence level (1.96), which means that if we reject 
the null hypothesis, there is greater than 5\% probability that 
we are wrong. On the contrary, the value $|z_1|=10.7$ for the SP03 data  
shows very bad agreement of the SAPHIR data with the model fit 2. 
Therefore, the hypothesis that fit 2 describes the SP03 data can be 
ruled out with a very high confidence. 
 
To estimate the relative global scaling factor between the CL05 
and SP03 data, we calculated the quantity 
\begin{equation}
\chi_0^2 = \sum_i \left(\frac{a\;\sigma_i^{\rm exp}-\sigma^{\rm th}
(E_i,\theta_i)}{\Delta\sigma_i^{\rm stat}}\right)^2 ,
\label{globalchi}
\end{equation}
using the SP03 data. The parameter $a$ was chosen to minimize 
$\chi_0^2$. For fit 1 and the full data set (sample A), $a=1.13$ 
and $\chi_0^2/N=4.80$. These values show that shifting the SP03 
data up by 13\% improves the agreement with the fit 1 model.  
For fit 2 and the forward-angle data (sample B), $a=1.15$ and 
$\chi_0^2/N=5.29$, which indicates 15\% scaling. These results 
are in good agreement with the estimated systematic uncertainties 
of the CL05 (8\%) and SP03 data. They are, however, smaller than 
the suggested scaling factor of $\approx 4/3$ \cite{CL05}. 
The coherent shift of the SP03 data with respect to the CL05, LEPS, 
and OLD ones is also apparent from the comparison 
of the appropriate values of $\langle R\rangle$ for fit 1 
(Table~\ref{sampleA}) and fit 2 (Table~\ref{sampleB}). Therefore, 
this analysis quantitatively shows that a combination of the CL05 
and SP03 data should not be considered in fixing the parameters 
of models, especially at forward angles. Instead, the use of 
the CL05, LEPS, and OLD data sets is the more preferred choice.

Refitting the fit 2 model parameters using the CL05, 
LEPS, and OLD data in the forward hemisphere 
($\theta_{\rm K}^{\rm c.m.} < 90^\circ$) yields 
$\chi^2 /{\rm n.d.f.} = 2.33$ and small changes in coupling constants. 
The largest changes appear for the coupling constants of the 
$s$-channel $D_{15}(1675)$ and $u$-channel $P_{01}(1810)$ resonances.
We note that 
the former is important for a proper description of the 
forward-angle and high-energy cross sections, which is necessary 
for fitting especially the LEPS data.   

Finally, let us discuss the physics consequence of the discrepancy 
between the CL05 and SP03 data on the fitted resonance parameters. 
As shown by the recent multipoles approach~\cite{Mart:2006dk}, the use
of these data sets individually or simultaneously leads to quite different 
parameters of resonances which, therefore, could lead to different 
conclusions about ``missing resonances''. 
Fitting to the SP03 data, e.g., indicates that the $S_{11}(1650)$, 
$P_{13}(1720)$, $D_{13}(1700)$, $D_{13}(2080)$, $F_{15}(1680)$, and 
$F_{15}(2000)$ resonances are required, while fitting to the CL05 
data leads alternatively to the $P_{13}(1900)$, $D_{13}(2080)$, 
$D_{15}(1675)$, $F_{15}(1680)$, and $F_{17}(1990)$ resonances.
Nevertheless, both  CL05 and SP03 support the existence of the 
missing $D_{13}(2080)$ resonance previously found in the Kaon-Maid 
model by using the SP98 data \cite{SP98} and denoted as 
$D_{13}(1895)$ (see Section~\ref{models}). It was found that the
extracted mass of this resonance would be 1936 (1915) MeV  
if the SP03 (CL05) data were used. We have refitted the original 
Kaon-Maid model to investigate this phenomenon. The result is shown 
in Table~\ref{tab:missing}. Obviously, the extracted values 
corroborate the finding of Ref.~\cite{Mart:2006dk}. The reason that 
the mass is slightly shifted to a higher value (as well as the broader 
width $\Gamma$ in the case of CL05) is obvious from the total cross 
section data (see the second peak of the total cross section 
shown in Fig.~9 of Ref.~\cite{Mart:2006dk}). 

\begin{table}[t]
  \centering
  \caption{Extracted values of mass $M$ and width $\Gamma$ of
    the missing $D_{13}$ resonance in Kaon-Maid using three different
    experimental data sets.}
  \label{tab:missing}
  \begin{ruledtabular}
  \begin{tabular}[c]{lccc}
    & Original (SP98)~\cite{SP98} & SP03~\cite{SP03} & CL05~\cite{CL05}\\
    \hline
    $M$ (GeV)      & $1.895\pm 0.004$ & $1.938\pm 0.004$ & $1.927\pm 0.003$\\
    $\Gamma$ (GeV) & $0.372\pm 0.029$ & $0.233\pm 0.008$ & $0.570\pm 0.019$\\
  \end{tabular}
  \end{ruledtabular}
\end{table}

\section{Conclusions}
\label{Conclusions}

We have analyzed the old (pre-1972) and new (CLAS 2005, SAPHIR 2003, 
and LEPS) experimental data by comparing them with several existing 
isobaric models, along with two new  models fitted to the CLAS data. 
Special attention was given to the forward-angles data, i.e. data 
with $\theta_K\leq 60^\circ$. The phenomenon of the cross section 
suppression at forward angles for the isobaric models with 
the hadronic form factors was observed.

At forward angles, the CLAS 2005, LEPS, and pre-1972 data can be 
described reasonably well within the isobaric model without hadronic 
form factors. The SAPHIR 2003 data are systematically shifted below  
the model predictions which requires a global scaling factor of 15\% 
to remove the discrepancy. The model without hadronic form factors, 
however, cannot describe the data in the backward hemisphere and 
at energies $W > 2$~GeV. 

The isobaric models with hadronic form factors were shown to give 
too strong damping of the cross sections at small kaon angles and 
energies $W > 1.9$~GeV, which results in a disagreement with 
existing experimental data. In their present forms, these models 
are therefore not suited for the description of photoproduction 
in this kinematic region, which is important, e.g., in the 
calculation of hypernuclear photoproduction.  Needless to say,  
more precise experimental data at very small kaon c.m. angles 
($0^\circ - 15^\circ$) would help solve this problem.

The Saclay-Lyon and Kaon-Maid models do not describe the data 
satisfactorily as indicated by the statistics $|z_1|$ for testing 
hypotheses. The former model is more consistent with the pre-1972 and 
LEPS data sets than with the CLAS 2005 and SAPHIR 2003 ones. 
At forward angles, the Saclay-Lyon model agrees quite well with 
the CLAS data. The Kaon-Maid model provides a better description 
of  the CLAS 2005, LEPS, and pre-1972 data than the SAPHIR 
2003 ones.

The relative-global-scaling factor between the SAPHIR and CLAS 
data is estimated to be 1.13, which is in agreement with the given 
systematic uncertainties. This discrepancy was shown to affect  
the parameters of the ``missing'' resonance $D_{13}(1895)$ in the 
Kaon-Maid model. The extracted values of the mass and width of the 
resonance differ by 11 and 337 MeV, respectively, when the SAPHIR 
and CLAS data are individually used in fitting the parameters. 
This finding agrees with the conclusion of a similar 
analysis that used the multipoles approach~\cite{Mart:2006dk}.  

\section{Acknowledgment}
The authors are grateful to O. Dragoun for useful discussions and 
interest in this work. P.B. acknowledges support provided by the 
Grant Agency of the Czech Republic, Grant No.202/05/2142 and the 
Institutional Research Plan AVOZ10480505. T.M. acknowledges the 
support from the Faculty of Mathematics and Sciences, UI, as well 
as from the Hibah Pascasarjana grant.

\renewcommand{\baselinestretch}{1.5}

\end{document}